\documentclass[11pt,a4paper,twoside,groupcitations]{article}
\usepackage[T1]{fontenc}
\usepackage[ansinew]{inputenc}
\usepackage[english]{babel}
\usepackage{amsfonts}
\usepackage{amsmath}
\usepackage{bm}
\usepackage{array}
\usepackage{amsthm}
\usepackage{amssymb}
\usepackage{graphicx}
\usepackage{braket}
\usepackage{eucal}
\usepackage{verbatim}
\usepackage[table]{xcolor}
\usepackage{caption}
\usepackage{cite}
\usepackage{textcomp}
\usepackage{float}
\usepackage{multicol}
\usepackage{tikz}
\usetikzlibrary{positioning,arrows}
\usetikzlibrary{decorations.pathmorphing}
\usetikzlibrary{decorations.markings}
\usetikzlibrary{calc,decorations.markings}
\usetikzlibrary{arrows,shapes}
\usetikzlibrary{matrix,arrows}

\usepackage[colorlinks,hyperindex,unicode]{hyperref}
\definecolor{green1}{RGB}{0,128,0} 
\hypersetup{hidelinks,backref=true,pagebackref=true,hyperindex=true,colorlinks=true,breaklinks=true,urlcolor= blue}
\hypersetup{%
  colorlinks = true,
  linkcolor  = blue,
  citecolor = cyan,
}
\newcommand{\beq}{\begin{eqnarray}}
\newcommand{\eeq}{\end{eqnarray}}
\newcommand{\be}{\begin{eqnarray}}
\newcommand{\ee}{\end{eqnarray}}
\newcommand{\pro}[2]{\mbox{$\langle\, #1 \mid #2\,\rangle$}}
\newcommand{\expec}[1]{\mbox{$\langle\, #1\,\rangle$}}

\renewcommand{\d}{\mbox{${\rm d}$}} 
\newcommand{\lp}{\ell_{\rm p}}
\newcommand{\mpl}{m_{\rm p}}
\newcommand{\gn}{G_{\rm N}}

\newcommand{\Rh}{R_{\rm H}}

\title{\bf Configurational entropy of black hole quantum cores}
\author{R.~Casadio$^{ab}$\thanks{E-mail: casadio@bo.infn.it},
$\ $
R.~da~Rocha$^{c}$\thanks{E-mail: roldao.rocha@ufabc.edu.br},
$\ $
P.~Meert$^{d}$\thanks{E-mail: pedro.meert@ufabc.edu.br},
$\ $
L.~Tabarroni$^{a}$\thanks{E-mail: luca.tabarroni@studio.unibo.it},
$ $
and
W.~Barreto$^{de}$\thanks{E-mail: willians.barreto@ufabc.edu.br}
\\
\\
$^a${\em Dipartimento di Fisica e Astronomia, Universit\`a di Bologna}
\\
{\em via Irnerio~46, 40126 Bologna, Italy}
\\
\\
$^b${\em I.N.F.N., Sezione di Bologna, I.S.~FLAG}
\\
{\em viale B.~Pichat~6/2, 40127 Bologna, Italy}
\\
\\
$^c${\em Federal University of ABC, Center of Mathematics}
\\
{\em Santo Andr\'e, 09210-580, Brazil.}
\\
\\
$^d${\em Federal University of ABC, Center of Natural Sciences}
\\
{\em Santo Andr\'e, 09210-580, Brazil}
\\
\\
$^e${\em Centro de Fisica Fundamental, Universidad de Los Andes,}
\\
{\em M\'erida~5101, Venezuela}
}
\begin{document}
\maketitle
\begin{abstract}
Two types of information entropy are studied for the quantum states of a model for the matter core inside
a black hole geometry.
A detailed description is first given of the quantum mechanical picture leading to a spectrum of bound states
for a collapsing ball of dust in general relativity with a non-trivial ground state.
Information entropies are then computed, shedding new light on the stability of the ground state
and the spectrum of higher excited states.
\end{abstract}
\section{Introduction}
\setcounter{equation}{0}
\label{S:intro}
Quantum aspects of gravitational collapse are among the most investigated topics in contemporary 
theoretical physics.
A quantum theory of gravity is expected to eliminate the singularities predicted by general relativity,
in particular, those associated with incomplete geodesics at the final stage of the collapse of regular matter
into a black hole~\cite{HE}.
Several methods for removing the singularity in approaches to quantum gravity have been
proposed~\cite{Hajicek:2001yd,Kuntz:2019lzq,Kuntz:2017pjd,Haggard:2014rza,Kuntz:2019gup,Piechocki:2020bfo}
and the appearance of a bounce at a minimum radius is generically obtained in semiclassical
models~\cite{frolov,Casadio:1998yr,Schmitz:2020vdr}.
These results suggest that the role played by matter in the description of black hole formation (and subsequent
evolution~\cite{Casadio:2019tfz}) is crucial. 
\par
The non-linearity of the Einstein equations makes it impossible to study any realistic model for
the gravitational collapse analytically, and this furthermore renders the problem of its quantum description
intractable in general.
For this reason, one can only go back to the study of (over)simplified models obtained by forcing a
strong symmetry and unphysical equation of state for the collapsing matter.
The prototypical example is given by the gravitational collapse of a ball (a perfectly spherical distribution)
of dust (matter with no other interaction but gravity) originally investigated by Oppenheimer and collaborators~\cite{OS}.
A similarly simple case is given by a shell of matter~\cite{mtw} collapsing under its weight or towards a central source.
These systems are simple enough to allow for a canonical analysis~\cite{kiefer} of the effective action obtained by restricting
the Einstein-Hilbert action to metrics satisfying the assumed symmetry properties.
\par
Here, we will carefully reconsider the spectrum of quantum states of the matter core inside a black hole found in Ref.~\cite{Casadio:2021cbv}
by studying the general relativistic model of gravitational collapse of a spherically symmetric ball of dust~\cite{OS,stephani}.
The key idea is to quantise the geodesic equation for the areal radius of the ball as an effective quantum mechanical description
of the dust ball similar to the usual quantum mechanical description of the hydrogen atom provided by the quantisation of
the electron's position.
This approach straightforwardly leads to the existence of a discrete spectrum of bound states.~\footnote{For similar results
for thin shells, see Ref.~\cite{vaz} (see also Refs.~\cite{hajicek,hk,Casadio:1995qy,Husain:2021ojz}).}
More importantly, one finds that the ball in the ground state properly compatible with general relativity has a quantised and
macroscopically large surface area, which resembles Bekenstein's area law~\cite{bekenstein}, so that no singularity
ever forms.~\footnote{The geometry sourced by this quantum core was reconstructed in Refs.~\cite{Casadio:2021eio}.}
Of course, the areal radius of the ball is not a fundamental degree of freedom for the matter in the collapsing core,
which should instead be described by quantum excitations of fields in the Standard Model of particle physics.
Although these fundamental degrees of freedom are neglected for the purpose of defining a tractable mathematical
problem, we expect that their existence should be encoded in a suitable entropy that can be computed from the states
of the effective quantum mechanical theory.
Obtaining an estimate of this entropy is precisely the main task of the present work.
\par
Several relevant measures of information entropy have been proposed, mainly in the last decade, and employed
to investigate gravitational systems and quantum field theories in the continuum limit.
Among them, the configurational entropy (CE)~\cite{Gleiser:2011di} and its differential version (DCE) were shown
to play a prominent role, since critical points of the CE and DCE usually correspond to preferred occupied states.
Given the configuration of a spatially-localised field, the CE measures the amount of information
it contains, as if that configuration were a message written in the alphabet represented by momentum space
modes, each mode contributing to the message with specific weighted probabilities.
In this description, each physical field configuration presents a distinct quantifiable signature of information entropy,
with associated shape complexity~\cite{Gleiser:2015aav}.
The DCE critical points correspond to those momentum space microstates occupied by the physical system
with higher configurational stability, hence pointing to the dominant physical states~\cite{Gleiser:2018jpd,Gleiser:2018kbq}.
In this context, complex physical systems do not only extremise the corresponding dynamical action but
they also tend to ``optimise'' the entropic information. 
Moreover, these information entropies measure the shape complexity encoded by the probability
density of a physical system.
The DCE is an effective mathematical tool for understanding pattern formation, and it has in fact been used
in a wide variety of cases, such as the study of the stability of AdS black holes~\cite{Braga:2016wzx},
the Hawking--Page transition regulated by a critical configurational instability~\cite{Braga:2019jqg},
brane-world models~\cite{Correa:2015vka}, neutron and boson
stars~\cite{Gleiser:2015rwa,Barreto:2022ohl,Lee:2017ero,Fernandes-Silva:2019fez,daRocha:2021jzn}.
The DCE provides a criterion for stability which was employed to compute the Chandrasekhar limit 
and critical stability domains for several stellar distributions~\cite{Gleiser:2013mga,Casadio:2016aum},
the spontaneous emission in hydrogen atoms~\cite{Gleiser:2017cro} and to study topological
defects~\cite{Gleiser:2020zaj,Gleiser:2019rvw,Barreto:2022mbx,Bazeia:2018uyg}, just to list a few examples.
\par
The main aim of this work is to explore the black hole quantum cores of Ref.~\cite{Casadio:2021cbv} and to discuss the important role
played by the different types of information entropy in extracting relevant knowledge about the spatial profile of the  quantum mechanical
system.
In this context, the information entropy and DCE will be shown to shed new light on the properties of the ground state and the spectrum
of higher modes as well.
Section~\ref{S:specttum} is devoted to presenting a refined version of the model for black hole quantum cores from
Ref.~\cite{Casadio:2021cbv}, with a derivation of the spectrum of bound states.
In Section~\ref{S:entropy}, the Shannon entropy~\cite{Shannon:1948zz} is introduced and the DCE and the information entropy
underlying the quantum mechanical description of black hole quantum cores are computed and analysed.
Section~\ref{S:conc} is dedicated to the concluding remarks. 
\section{Core quantum spectrum}
\setcounter{equation}{0}
\label{S:specttum}
As we recalled in the Introduction, the Oppenheimer-Snyder model is simple enough to allow for a
rigorous canonical analysis~\cite{Piechocki:2020bfo}.
However, some key features are more easily obtained by following a simpler approach introduced in Ref.~\cite{Casadio:2021cbv},
which we here review and improve upon.
\par  
Let us consider a perfectly isotropic ball of dust with total ADM mass~\cite{adm} $M$ and areal radius $R=R(\tau)$,
where $\tau$ is the proper time measured by a clock comoving with the dust. 
Dust particles inside this collapsing ball will follow radial geodesics $r=r(\tau)$ in the Schwarzschild space-time
metric~\footnote{We shall always use units with $c=1$ and often write the Planck constant $\hbar=\lp\,\mpl$ and
the Newton constant $\gn=\lp/\mpl$, where $\lp$ and $\mpl$ are the Planck length and mass, respectively.}
\be
\d s^{2}
=
-\left(1-\frac{2\,\gn\,M_0}{r}\right)
\d t^{2}
+\left(1-\frac{2\,\gn\,M_0}{r}\right)^{-1}
\d r^{2}
+r^{2}
\left(\d\theta^{2}+\sin^{2}\theta\, \d\phi^{2}\right)
\ ,
\label{schw}
\ee
where $M_0=M_0(r)$ is the (constant) fraction of ADM mass inside the sphere of radius $r=r(\tau)$.
\par
In particular, we can consider the outermost (thin) layer of (average) radius $r=R(\tau)$ and mass $\mu=\epsilon\,M$,
where $0<\epsilon<1$ is the fraction of the ball mass in the chosen layer.
The equation governing the evolution of the radius of this layer is then given by the mass-shell condition for the massive layer
of four-velocity $u^\alpha=(\dot t,\dot R,0,0)$, that is
\be
\label{geod-general}
\frac{E_{\mu}^{2}}{\mu^{2}}
-
\dot R^{2}
+
\frac{2\,\gn\,M_{0}}{R}
-
\left(1-\frac{2\,\gn\,M_{0}}{R}\right)
\frac{L_{\mu}^{2}}{R^{2}\,\mu^{2}}
=
1
\ ,
\ee
where $M_0=(1-\epsilon)\,M$, and $E_\mu$ and $L_\mu$ are the conserved momenta conjugated to $t=t(\tau)$ and $\phi=\phi(\tau)$,
respectively.
For the particular case of radial motion, one must set $L_\mu=0$ and Eq.~\eqref{geod-general} reads
\be
\label{geod-part}
H
\equiv
\frac{P^{2}}{2\,\epsilon\, M}
-\frac{\epsilon\left(1-\epsilon\right)\gn\,M^{2}}{R}
=
\frac{\epsilon\, M}{2}\left(\frac{E_\mu^{2}}{\epsilon^2\,M^{2}}-1\right)
\equiv
\mathcal E
\ ,
\ee
where $P=\mu\,\dot R$ is the momentum conjugated to $R=R(\tau)$.
Notice that Eq.~\eqref{geod-part} contains the parameter $\epsilon$, and all of the results will therefore depend on the distribution
of dust between the outermost layer of mass $\mu=\epsilon\,M$ and the inner spherical core of mass $M_0=(1-\epsilon)\,M$.
This represents a first improvement concerning the more qualitative analysis of Ref.~\cite{Casadio:2021cbv}.
\par
Having established the equation of motion governing the collapse, we apply the canonical quantization prescription 
\be
P\mapsto\hat{P}
=-i\,\hbar\,\partial_R
\ ,
\ee
which allows us to write Eq.~\eqref{geod-part} as the time-independent Schr\"odinger equation
\be
\hat{H}\,\Psi_{\bar n}
=
\left[
-
\frac{\hbar^{2}}{2\,\epsilon\, M}
\left(
\frac{\d^2}{\d R^2}
+
\frac{2}{R}\,
\frac{\d}{\d R}
\right)
-
\frac{\epsilon\left(1-\epsilon\right)\gn\,M^{2}}{R}
\right]
\Psi_{\bar n}
=
\mathcal E_{\bar n}\,
\Psi_n
\ .
\ee
The above equation is formally the same as the one for the $s$-states of the hydrogen atom,~\footnote{Perfect isotropy implies
that the angular momentum quantum numbers $l=m=0$.
The case of a (slowly and rigidly) rotating ball is left for future development.}
so that the solutions are given by the Hamiltonian eigenfunctions
\be
\label{radial-wavefunction}
\Psi_{\bar{n}}(R)
=
\sqrt{\frac{\epsilon^{6}\left(1-\epsilon\right)^{3}M^{9}}{\pi\,\lp^{3}\,\mpl^{9}\,\bar{n}^{5}}}\,
\exp\!\left(-\frac{\epsilon^{2}\,(1-\epsilon)\,M^{3}\,R}{\bar{n}\,\mpl^{3}\,\lp}\right)
L_{\bar{n}-1}^{1}\!\!
\left(\frac{2\,\epsilon^{2}\,(1-\epsilon)\,M^{3}\,R}{\bar{n}\,\mpl^{3}\,\lp}\right)
\ ,
\ee
where $L_{{\bar n}-1}^1$ are Laguerre polynomials and ${\bar n}=1,2\,\ldots$, corresponding to the eigenvalues 
\be
\mathcal E_{\bar n}
=
-
\frac{\epsilon^3\,(1-\epsilon)^2\,M^5}{2\,\mpl^4\,\bar n^2}
\ .
\ee
Note that the normalisation is defined in the scalar product which makes the Hamiltonian $\hat H$ Hermitian 
when acting on the above spectrum, that is
\be
\pro{\bar n}{\bar n'}
=
4\,\pi\,\int_0^\infty
R^2\,\Psi_{\bar n}^*(R)\,\Psi_{\bar n'}(R)\,
\d R
=
\delta_{\bar n \bar n'}
\ .
\ee
The expectation value of the areal radius on these eigenstates is thus given by
\be
\bar R_{\bar n}
\equiv
\bra{\bar n} \hat R \ket{\bar n}
=
\frac{3\,\mpl^3\,\lp\,\bar n^2}{2\,\epsilon^2\,(1-\epsilon)\,M^3}
\ .
\ee
As was noted in Ref.~\cite{Casadio:2021cbv}, so far the quantum picture is the same that one would have in Newtonian
physics, with the ground state $\bar n=1$ having a width $\bar R_1\sim \lp\,(\mpl/M)^3$ and energy $\mathcal E_1\sim -M\,(M/\mpl)^4$.
This state is practically indistinguishable from a point-like singularity for any macroscopic black hole of mass $M\gg\mpl$.
\par
In fact, the only general relativistic feature that the model retains is given by the form of $\mathcal E=\mathcal E(E_\mu)$
in Eq.~\eqref{geod-part}.
By then assuming that the conserved $E_\mu$ remains well-defined for the allowed quantum states, we obtain
\be
\label{Emu}
0
\le
\frac{E_\mu^2}{\epsilon^2\,M^2}
=
1
-
\frac{\epsilon^2\,(1-\epsilon)^2}{n^2}
\left(\frac{M}{\mpl}\right)^4
\ ,
\ee
which yields the lower bound for the principal quantum number~\footnote{A similar quantisation for the mass $M$ was
found long ago in Refs.~\cite{Kaup:1968zz,Ruffini:1969qy} by studying stable configurations of boson stars.
As we recalled in the Introduction, Eq.~\eqref{N_M} is also reminiscent of Bekenstein's area law~\cite{bekenstein}.}
\be
\bar n
\ge
N_M
\equiv
\epsilon\,(1-\epsilon)
\left(\frac{M}{\mpl}\right)^2
\ .
\label{N_M}
\ee
Saturating the inequality \eqref{N_M} corresponds to the fundamental state of the outer layer, which results in a huge number
of states (with $\bar n<N_M$) that remain inaccessible for the collapsing ball.
Among these states, there is also the one discussed above with $\bar R_{\bar n=1}\sim 0$, which means that the singularity
is then precluded, as one expects from semiclassical models~\cite{frolov,Casadio:1998yr}.
More precisely, since the only information that one can extract from the wave equation, and confront with a classical quantity,
is the areal radius $\expec{\hat R}$, one does not identify the potential singularity with a principal quantum number $\bar{n}$,
rather than using the value of $R$ corresponding to that $\bar{n}$.
All quantum states can access $R = 0$ \emph{a priori}, however, since the lower bound for $\bar{n}$ is much bigger than
the one corresponding to  $R\sim \bar R_1\sim 0$, the probability that the shell of dust is found in the singularity
is practically zero for $M\gg\mpl$.
\par
We recall that there are no reasons for a ball of dust evolving solely under the gravitational force, to stop contracting
classically.
Quantum mechanics instead shows that, in order to have a well-defined energy spectrum, the proper ground state compatible
with general relativity has a width
\be
\bar R_{N_M}
=
\frac{3}{4}\,(1-\epsilon)\,\Rh
\ ,
\ee
where $\Rh=2\,\gn\,M$ is the classical Schwarzschild radius of the ball.
Since $0<\epsilon<1$, one can conclude that $\bar R_{N_M}<\Rh$ and a finite number of states~\eqref{radial-wavefunction}
with 
\be
\bar n=N_M+n
\label{ndef}
\ee
and $n=0,1,\ldots$ will exist inside the horizon.
\par
In the above analysis, the parameter $\epsilon$ remains undetermined.
One of the goals of this work is to determine a preferred value from entropic considerations.
Before we do that, we notice that the uncertainty in the areal radius is given by
\be
\frac{\Delta R_{\bar n}}{\bar R_{\bar n}}
\equiv
\frac{\sqrt{\bra{\bar n}\hat R^2\ket{\bar n}-\bar R_n^2}}{\bar R_n}
=
\frac{\sqrt{\bar n^2+2}}{3\,\bar n}
\ ,
\ee
which asymptotes to a minimum of $1/3$ for $\bar n\to \infty$.
We thus expect that $N_M\gg 1$  for the ground state and consequently $\epsilon\,(1-\epsilon)\sim 1$.
For instance, for $M=M_\odot\simeq 10^{30}\,$kg, one finds $N_M\sim 10^{76}$, which makes it impossible
to handle the wavefuntion~\eqref{radial-wavefunction}, either analytically or numerically.
We shall therefore limit our investigation to values of $M$ small enough to allow for simple numerical calculations
and extrapolate to larger values of $M$.
\section{Configurational Entropy}
\setcounter{equation}{0}
%
\label{S:entropy}
The CE is a measure of the amount of information carried by physical solutions to the equations of
motion of a given theory~\cite{Gleiser:2011di}.
A way to come up with it is by considering the original formulation of information entropy by
Shannon~\cite{Shannon:1948zz}:
information is measured by the minimum number of bits needed to convert symbols into a coded form,
in such a way that the highest transmission rate of messages, consisting of an array of the given symbols, 
can be communicated through a channel.
Shannon then defined a measure of information by the expression
\be
S_{\rm Sh} = -\sum_{c_i\in{\mathcal{C}}}p_{c_i}\log p_{c_i}
\ ,
\label{Ssh}
\ee
where the sum is taken over the set $\mathcal{C} = \{c_i;\ i=1,\ldots,N\}$ of symbols with a distribution $p_i = p(c_i)$
determining their occurrence.
The amount of information carried by each symbol was determined by Shannon as $\mathbb{I}(c_i) = -\log_2 p_i$,
so that
\be
\label{ShannonInfo}
\expec{ \mathbb{I}}
=
\sum_{c_i\in\mathcal C} p(c_i)\, \mathbb{I}(c_i)
=
-
\sum_{c_i\in\mathcal C}
p(c_i)\, \log_2 p(c_i)
=
S_{\rm Sh}
\ .
\ee
In fact, $\log_2 N$ is the number of bits to describe $N$ symbols and the entropy is the highest
possible for a uniform distribution with $p(c_i) = 1/N$, that is $\expec{ \mathbb{I} } = \log_2 N$.~\footnote{Note that
$S_{\rm Sh}$ is dimensionless and the base of the logarithm in Eq.~\eqref{Ssh} is a matter of choice.
The binary base 2 measures the information entropy in bits (or Sh for Shannon), but one can also employ
the natural logarithm to measure the information entropy in the natural units called nat, with $1\,$nat$=$$(1/\ln 2)\,$Sh.}
\par
Shannon entropy can be straightforwardly connected with the Gibbs entropy $S_{\rm G}$, which generalises the Boltzmann
entropy $S_{\rm B}$ to thermodynamic systems described by microstates that do not necessarily have equal
probabilities, that is
\be
S_{\rm G}
=
-k_{\rm B}\,
\sum_a p_{a}\,\ln p_{a}
\ ,
\label{Sg}
\ee
where $k_{\rm B}$ is the Boltzmann constant and $p_a$ stands for the probability of occurrence of each
microscopic configuration $a$ of the system.
Besides the overall constant $k_{\rm B}$, which is needed to give the entropy its natural thermodynamical units,
one finds $S_{\rm G}\sim S_{\rm Sh}$
if one can identify $p_a=p(c_i)$ for a given thermodynamical system.
Moreover, when $p_a=p$ for all microstates in the given system, $S_{\rm G} = S_{\rm B}=k_{\rm B}\,\ln(W)$,
where $W$ is the number of equiprobable microstates.
\par
The CE is a form of Shannon entropy for static macroscopic systems computed by assuming that the underlying microstates
$a$ in $S_{\rm G}$ are defined by plane or standing waves of wavenumber $\vec k$ in flat momentum space, that is
\beq
a_{\vec k}=e^{i\,\vec k\cdot\vec x}
\ .
\label{ak}
\eeq
In order to compute the probability $p({\vec k})$ of occurrence of each microstate $a_{\vec k}$, one can therefore start
by taking the Fourier transform $\rho(\vec k)$ of the probability density $\rho(\vec x)$ of a given macroscopic system
in position space.
The probability $p(\vec k)$ is then defined by a suitable normalisation such that $\sum_{\vec k} p(\vec k)=1$.
We also recall that the power spectrum $\sum_{\vec k} |\rho(\vec{k})|^2 e^{i\,\vec k\cdot\vec x}$ is the Fourier transform
of the 2-point correlation function and encodes the shape complexity of the probability density.
\par
To introduce the DCE for the spherically symmetric wavefunctions in Eq.~\eqref{radial-wavefunction},
we start from the probability density 
\beq
\rho(R)
=
4\,\pi\, R^2\,\left|\Psi_{n}(R)\right|^2
\ ,
\label{lambda}
\eeq
with $n\ge 0$ defined in Eq.~\eqref{ndef}.
The spectral Fourier transform of Eq.~\eqref{lambda}, in spherical coordinates, can be written as
\begin{equation}
\rho({k})
=
\frac{1}{k^{1/2} }
\int_{0}^\infty 
R^{{3}/{2}}\,
\rho(R)\,
J_{1/2}(k\,R)\,
\d R
\ ,
\label{sphFourier}
\end{equation}
where $J_{1/2}$ denotes the Bessel function of order $1/2$. 
In the numerical calculations, an equivalent Hankel transform is implemented, which is proportional
to the radially symmetric spherical Fourier transform by a  factor of $(2\pi)^{3/2}$.
In the computations that follow, the function $\rho(k)$ is therefore given by the Hankel transform of the probability
density~\eqref{lambda} associated with our wavefunctions. 
\par
From Eq.~\eqref{sphFourier}, one can immediately define the so-called information entropy as the continuous version
of the Shannon entropy in Eq.~\eqref{Ssh}, which reads
\beq
\label{IE}
S_{\rm Sh}
=
-\int_0^\infty
\rho(k) \ln\left[\rho(k) \right]
\d k
\ .
\eeq 
The continuum limit can be problematic, and alternative expressions for the entropy have thus been conjectured.
\par
Gleiser and Sowinski introduced the DCE by first defining the modal fraction~\cite{Gleiser:2018jpd,Gleiser:2018kbq}
as a measure for the relative weight of a given mode in the probability distribution.
In our case, the modal fraction is explicitly given by
\begin{equation}
\label{modalfraction}
\tilde{\lambda}(k)
=
\frac{\left|\rho(k)\right|^{2}}
{\int_0^\infty \left|\rho(k')\right|^{2}\,\d k'}
\ 
\end{equation}
which measures the weight of each wave mode contributing to the total power spectrum.~\footnote{Indeed,
the DCE attains its maximum value for a uniform power spectrum.}
The DCE measures the number of bits that are necessary to describe localised physical configurations,
whose underlying information has the best compression and maximal channel capacity.
It can be defined as the functional
\begin{equation}
S_{\rm DCE}[\tilde{\lambda}]
=
-\int_0^\infty
\hat{\lambda}(k)\,
\ln\!\left[\hat{\lambda}(k)\right]\,
\d k
\ ,
\label{dce11}
\end{equation}
with~\footnote{This last step is introduced to ensures that $\hat{\lambda}(k)\leq1$ for all values of $k>0$.}
\beq
\hat{\lambda}(k)
=
\frac{\tilde{\lambda}(k)}
{\tilde{\lambda}_{\rm MAX}}
\ ,
\eeq
where $\tilde{\lambda}_{\rm MAX}$ is the peak of the momentum distribution $\tilde\lambda=\tilde\lambda(k)$.
The DCE can therefore be used to estimate the dynamical degree of order regulating the spectral density~\eqref{sphFourier}
and the configurational stability of the black hole quantum core will be determined by the critical points of the DCE.
The higher the DCE, the more delocalised the density, the higher the uncertainty, and the smaller the accuracy
in predicting the spatial localisation of the system.
One might also speculate that there is a relationship between the DCE and the effect of tracing out degrees of freedom
in the present context. 
\par
\begin{figure}[t]
\centering
\includegraphics[width=0.48\linewidth]{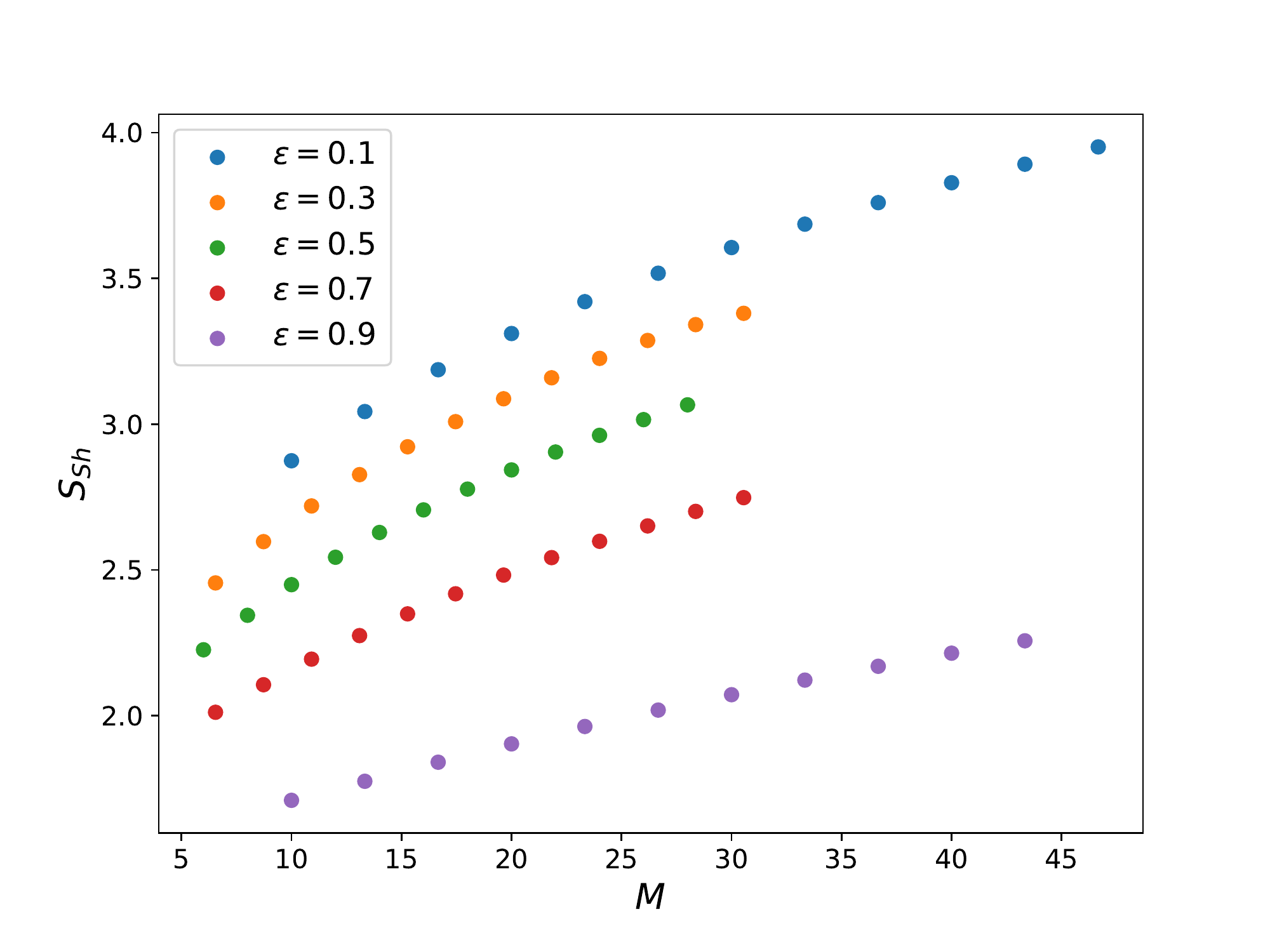}
\includegraphics[width=0.48\linewidth]{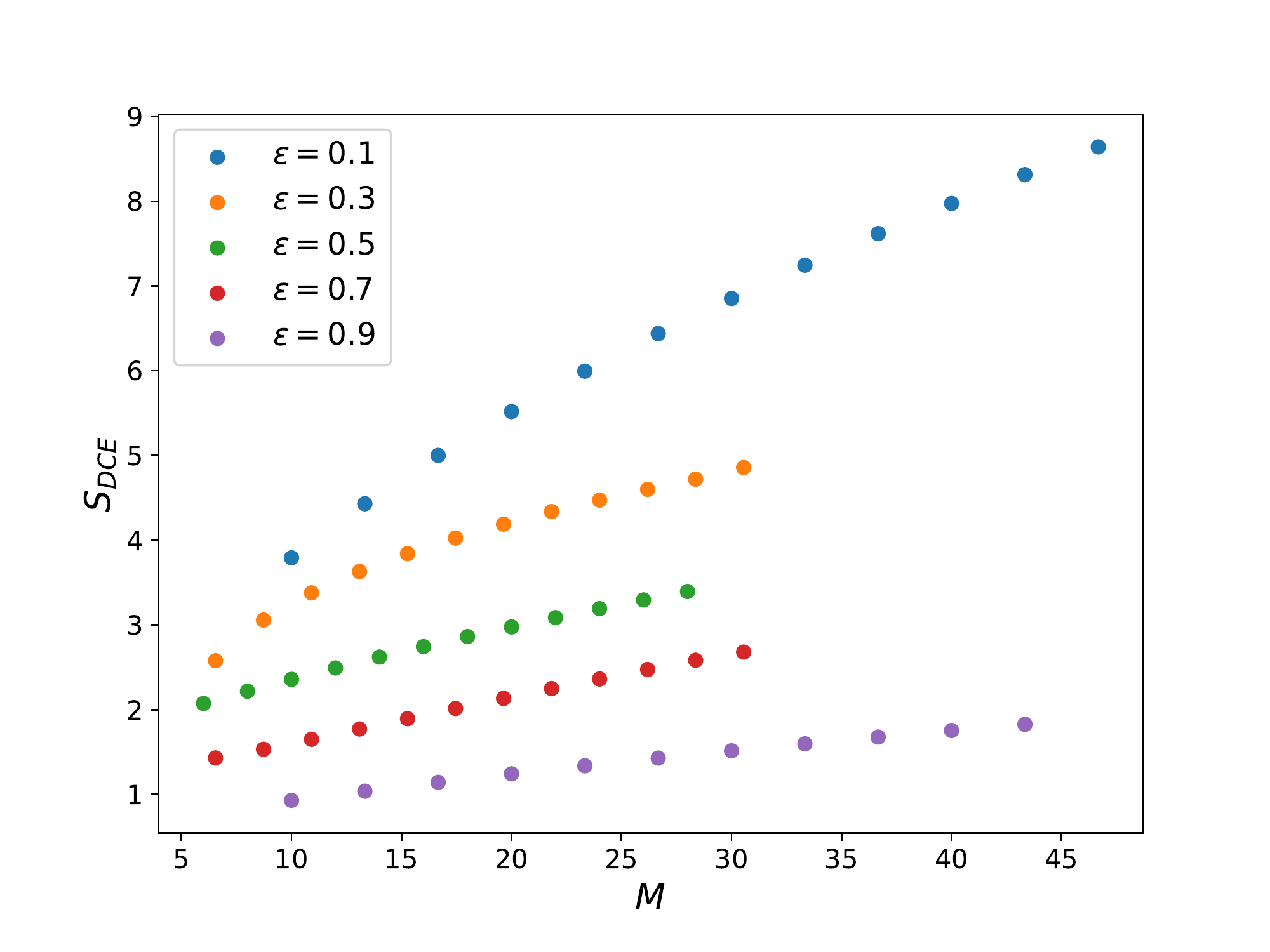}
\caption{Information entropy~\eqref{IE} (left panel) and DCE (right panel) for ground states ($n=0$)
of different mass $M$ and fixed $\epsilon$.
Note the different scales on the vertical axes (mass and entropies in Planck units).}
\label{fig:GS}
\end{figure}

We first computed numerically the Fourier transform~\eqref{sphFourier} and the modal fraction~\eqref{modalfraction}
for the ground state with $n=0$.
The resulting DCE is displayed in Fig.~\ref{fig:GS}, where we include the information entropy for comparison,
and all quantities are expressed in units of $\lp$ and $\mpl$ (equivalent to setting $\hbar=\gn=1$).
The left panel of Fig.~\ref{fig:GS} shows that the information entropy $S_{\rm Sh}$ is a monotonic increasing
function of the mass $M$, for each fixed value of the parameter $\epsilon$, and a monotonic decreasing function
of $\epsilon$, for a fixed value of $M$.
A similar profile holds for $S_{\rm DCE}$, as shown in the right panel.
In absolute values, comparing the two panels of Fig.~\ref{fig:GS} makes explicit the subtle distinction between the
information entropy~\eqref{IE} and the DCE~\eqref{dce11}.
Although both entropies have similar qualitative behaviour, their values are different for the same $M$ and $\epsilon$.
We recall that the outermost thin layer has (average) radius $R$ and mass $\mu = \epsilon\,M$, with $0 < \epsilon < 1$,
so that $\epsilon$ is the fraction of the ball mass in the chosen layer.
From Fig.~\ref{fig:GS}, we can then conclude that the higher the amount of dust in the outer layer,
the lower both the information entropy and the DCE. 
\par
\begin{figure}[t]
\centering
\includegraphics[width=0.48\linewidth]{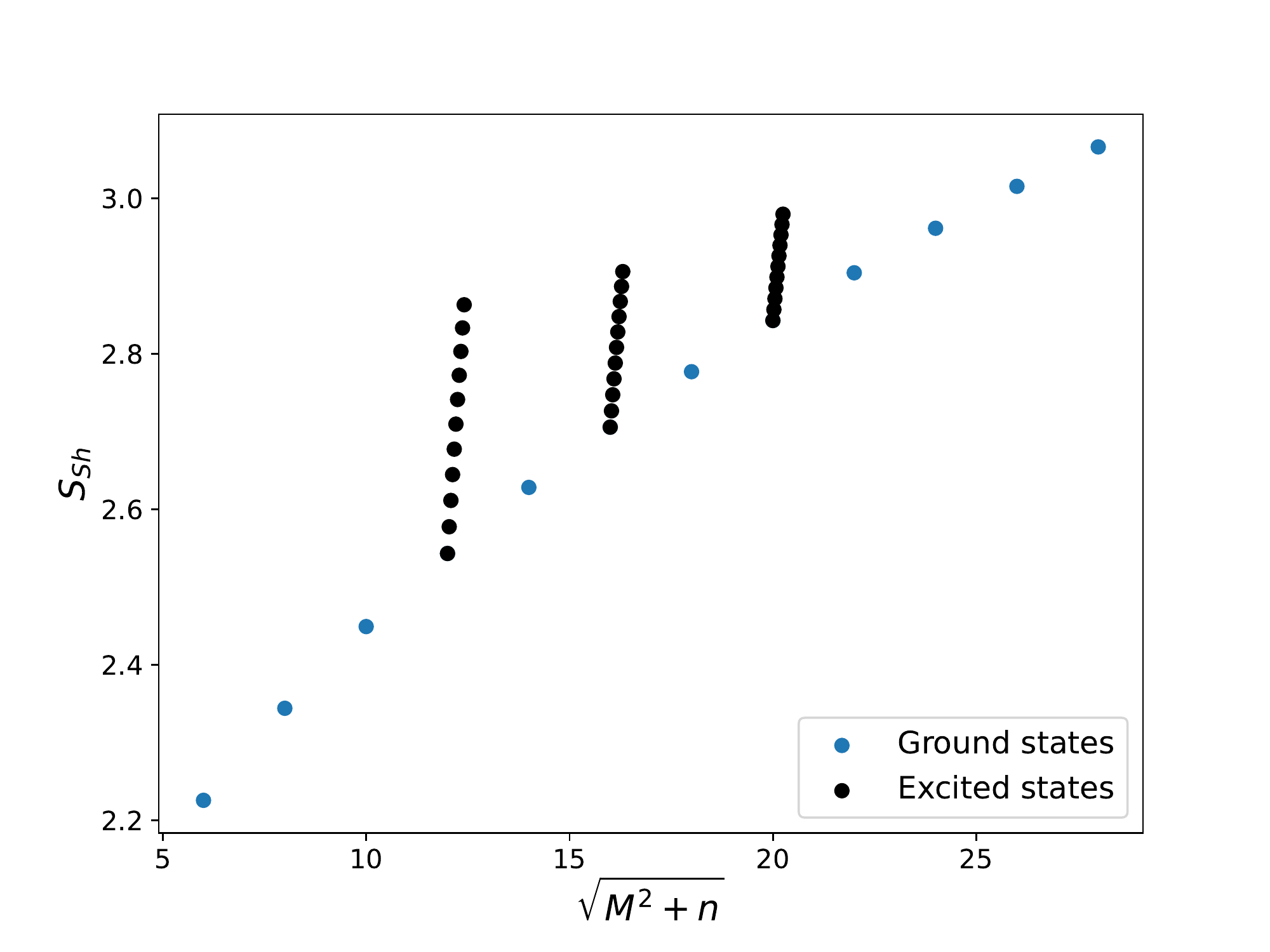}
\includegraphics[width=0.48\linewidth]{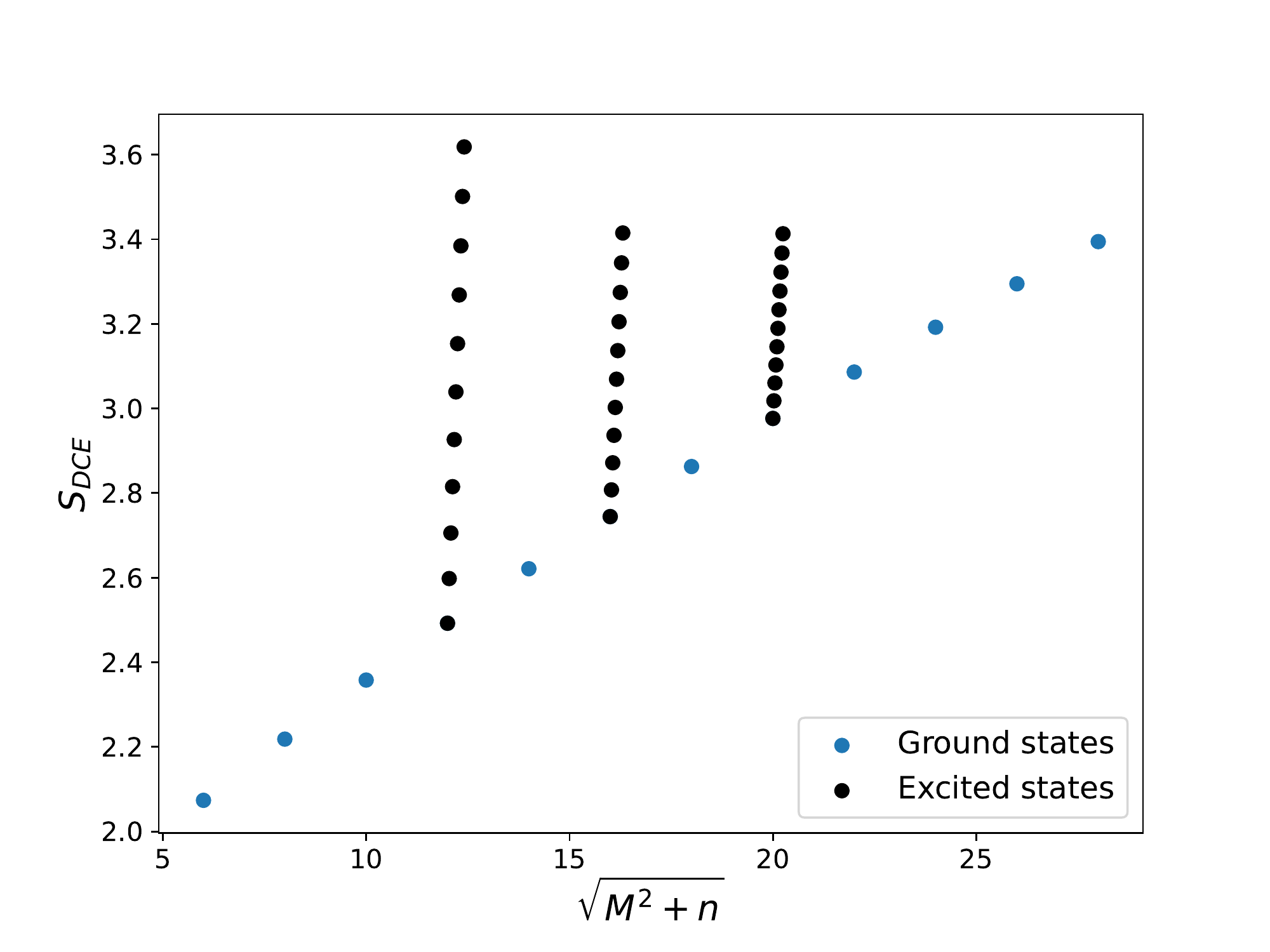}
\caption{Information entropy (left panel) and DCE (right panel) for  ground ($n=0$) and excited ($n>0$) states
with $\epsilon=0.5$ (mass and entropies in Planck units). }
\label{55}
\end{figure}

We next computed the information entropy and the DCE for the first few excited states with $n> 0$.
The results for $M=6$, $8$ and $10$ are shown in Figs.~\ref{55}, along with the entropies of the
corresponding ground states, for $\epsilon=0.5$.
(The branching out from the ground states is similar for other values of $\epsilon$.)
We remark that all of the values of the entropy are plotted as discrete points, since the
Laguerre polynomials in the wavefunction~\eqref{radial-wavefunction} must be of integer
order (the spectrum of allowed states is discrete).
We also notice that the plots show a range of very small values of the mass $M$, in units of the Planck mass,
because the order of the Laguerre polynomials growing with $M^2$ makes it numerically impossible to
reach masses of astrophysical relevance.
\par 
The main result clearly shown in Figs.~\ref{55} is that excited states have higher information entropy 
and DCE than the respective ground state.
Both kinds of entropy steeply increase, which supports the fact that excited states are unstable and 
will naturally decay into the ground state.
It is a general rule of the continuous limit of the Shannon entropy that states with higher information entropy
are less stable and can eventually be less predominant from the phenomenological point of view.  
Moreover, excited quantum states have similarities with the ground state in Fig.~\ref{fig:GS},
since higher values of $\epsilon$
correspond to lower $S_{\rm Sh}$ and $S_{\rm DCE}$, for each fixed value of $\bar{n}$.
\begin{figure}
\centering
\includegraphics[width=0.48\linewidth]{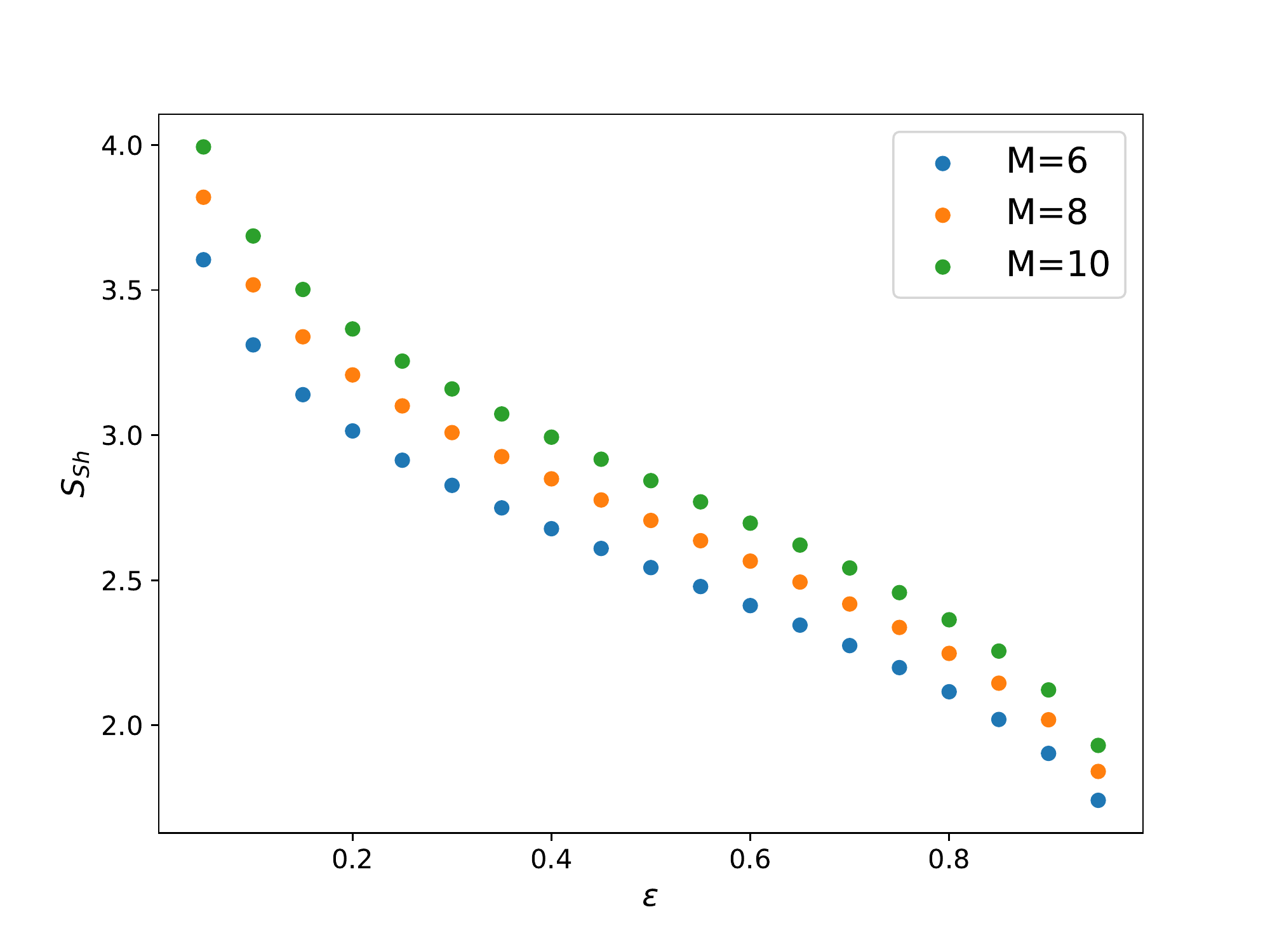}
\includegraphics[width=0.48\linewidth]{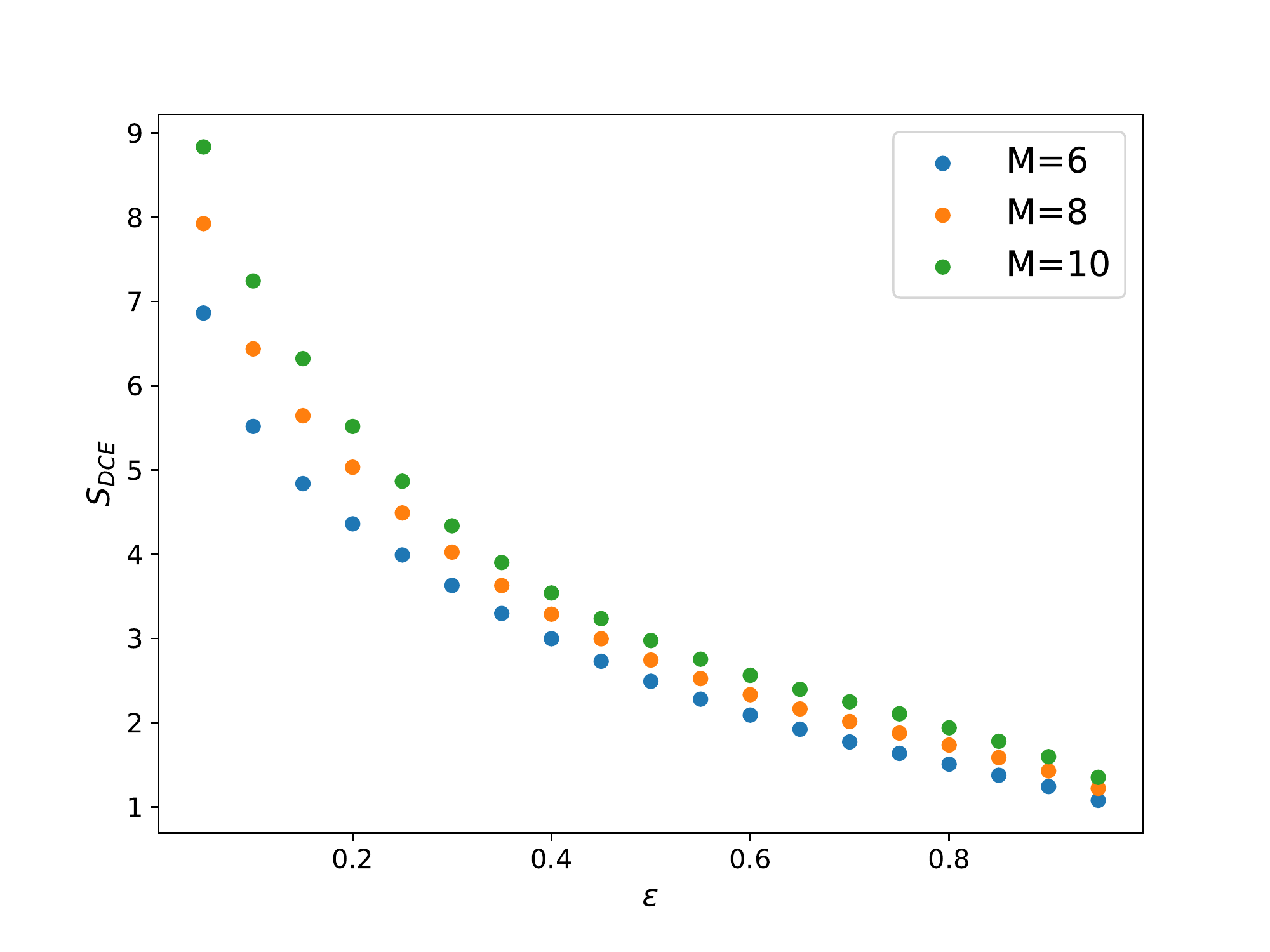}
\caption{Information entropy (left panel) and DCE (right panel) for the ground states ($n=0$) with
varying $\epsilon$ at fixed mass (mass and entropies in Planck units).}
\label{66}
\end{figure}
\par
It is also interesting to look at the relation between the entropy and the fraction $\epsilon$ of mass in the 
outermost layer for fixed values of the total mass $M$.
For this purpose, it is computationally more convenient to employ the rescaling of parameters described
in Appendix~\ref{A:rescaling}, and some results are then displayed in Fig.~\ref{66}.
It is clear that both Shannon's entropy and the DCE decrease as the outermost layer contains relatively more mass.
This is in agreement with the idea that the (information) entropy measures our ignorance of the microstates
of the system:
the larger $\epsilon$, the less mass is in the interior and the less information is missed by just considering
the state of the outermost layer.
\par
\begin{figure}[t]
\centering
\includegraphics[width=0.48\linewidth]{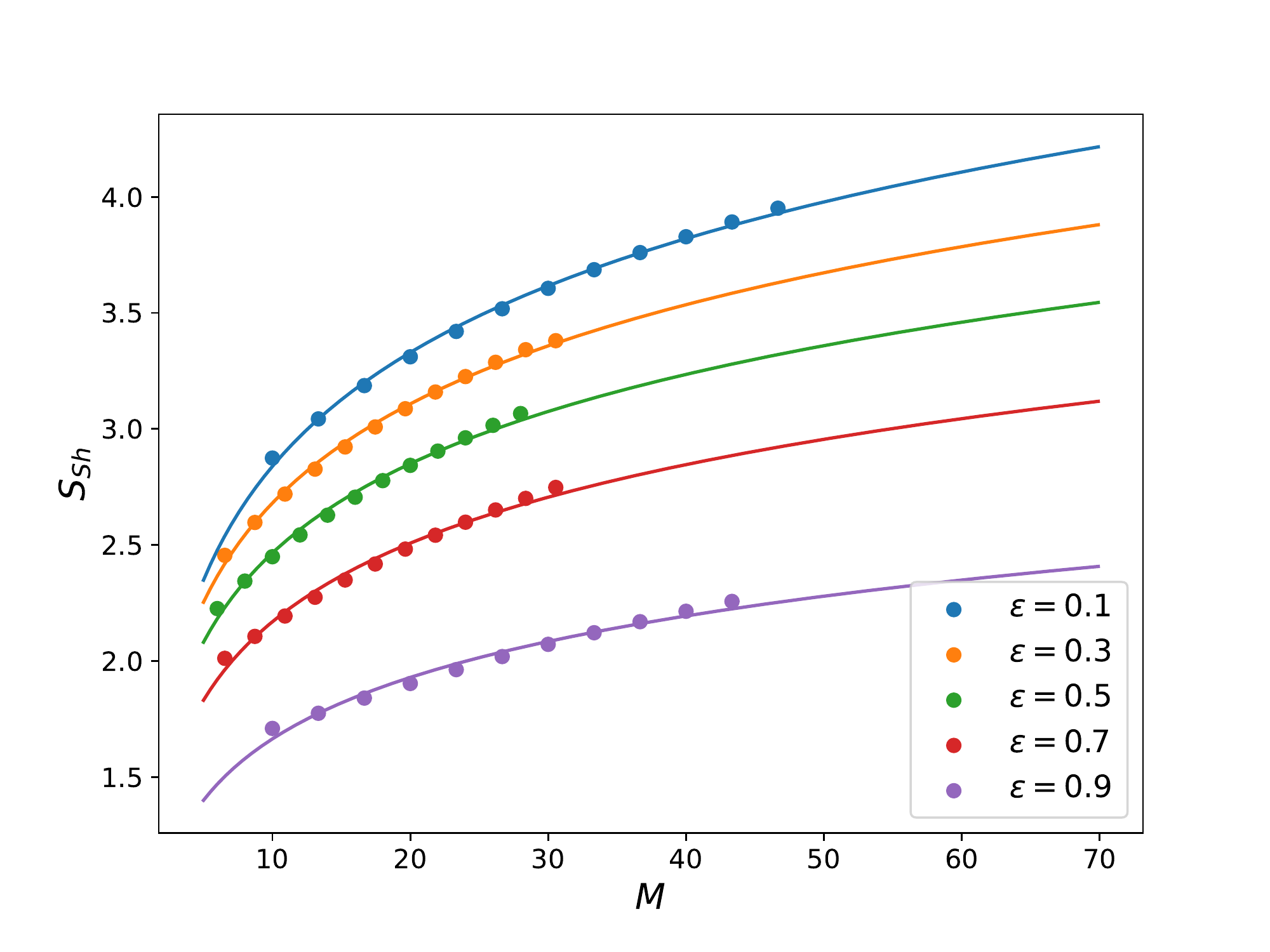}
\includegraphics[width=0.48\linewidth]{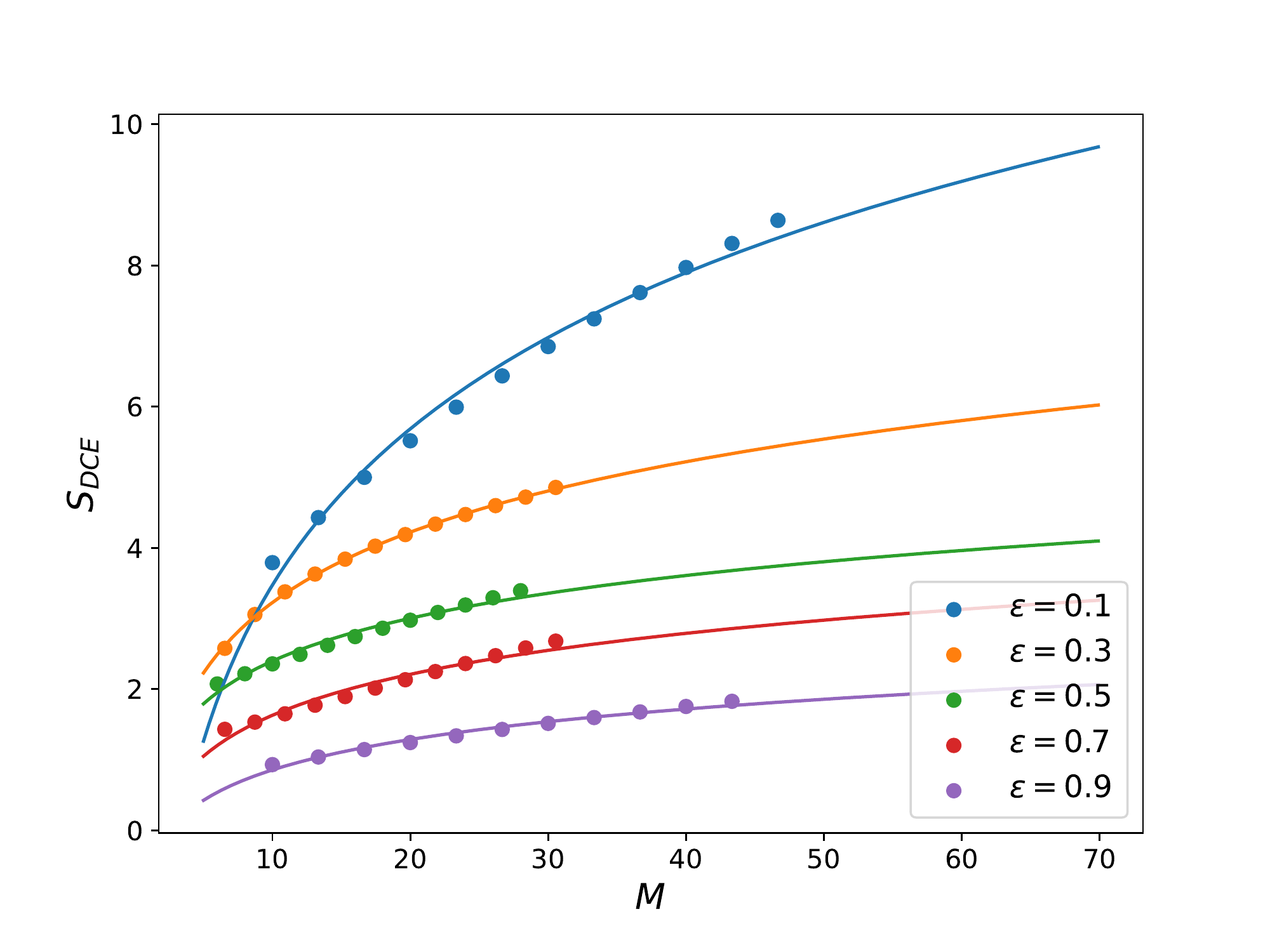}
\caption{Fitting curves for the information entropy (left panel) and DCE (right panel) for the ground states ($n=0$,
mass and entropies in Planck units).}
\label{fig:fit}
\end{figure}

Finally, we look back at the ground state entropies and argue whether Shannon's information entropy
and the DCE in Fig.~\ref{fig:GS} follow specific scaling laws.
In fact, we can reproduce quite accurately the numerically computed values with several interpolating curves,
among which logarithmic ones show the best fit, given by
\be
S
=
\alpha\,\ln\left( \frac{M}{\mpl}\right)+\beta
\ ,
\label{fit:eq}
\ee
where the fitting parameters for each curve are displayed in Table~\ref{table:SE}.
The fitting curves and their extrapolation to larger values of $M$ are displayed in Fig.~\ref{fig:fit}.
The coefficients of determination~\footnote{We recall that $R^2\simeq 1$ for a good agreement between data and the model.}
for the fitting model~\eqref{fit:eq} with the parameters in Table~\ref{table:SE} are all given by $R^2\approx 0.99$ for the information
entropy and $R^2=0.98\pm 0.01$ for the DCE. 
From Fig.~\ref{fig:fit}, it looks like $S_{\rm DCE}\sim M$ for $\epsilon=0.1$, and for such small values of $\epsilon$ the $R^2$ test
returns similar values for both fits.
However, for larger values of $\epsilon$, the linear regressions yields a worse $R^2\approx0.95$.
This suggests that the apparent linear behaviour of the DCE for small $\epsilon$ is just the approximation of a logarithm
small $M$ (as we already mentioned, computing the entropies for large $M$ is computationally very intense). 
The stability of the $R^2$ test for all the curves is a strong indication that Eq.~\eqref{fit:eq} is a
good approximation for our results.
\begin{table}[t]
	\centering
\begin{tabular}{|c|c|c|c|c|c|}
	\hline
&	$\epsilon=0.1$	& $\epsilon=0.3$ & $\epsilon=0.5$ & $\epsilon=0.7$ & $\epsilon=0.9$ \\
	\hline\hline
$S_{\rm Sh}$ &	$\alpha=0.707$	& $\alpha=0.616$ & $\alpha=0.554$ & $\alpha=0.488$ & $\alpha=0.382$ \\
	\hline
$S_{\rm Sh}$ &	$\beta=1.209$	& $\beta=1.260$ & $\beta=1.187$  & $\beta=1.044$ & $\beta=1.044$\\
	\hline\hline
$S_{\rm DCE}$ &	$\alpha=3.190$	& $\alpha=1.438$ & $\alpha=0.874$ & $\alpha=0.837$ & $\alpha=0.621$\\
		\hline
$S_{\rm DCE}$ &		$\beta=-3.8717$	& $\beta=-0.085$ & $\beta=0.385$ & $\beta=-0.297$ & $\beta=-0.297$ \\
		\hline
\end{tabular}
\caption{Fitting parameters for information entropy and DCE in the ground states ($n=0$).}
\label{table:SE}
\end{table}
\par 
Since the mass $M$ is proportional to the volume of the matter core, the above scaling relations are compatible
with the interpretation that the number of microscopic degrees of freedom counted by the exponential of the entropy
grows with the volume of the ball to some power $0<\alpha<1$. 
This is indeed the expected behaviour for matter and should be contrasted with the Bekenstein area law~\cite{bekenstein},
according to which the number of gravitons involved in the geometry sourced by this core~\cite{Casadio:2021eio}
scales like the horizon area, which is instead proportional to $M^2$ in Planck units.
Moreover, our results corroborate the fact that the DCE is an estimator for the number of bits one needs
to reconstruct the quantum-mechanical field configuration from the wave modes in momentum space.  
The matter entropies computed in this work, therefore, remain sub-dominant with respect to the gravitational
contribution for a (quantum) black hole. 
\section{Concluding remarks and discussion}
\setcounter{equation}{0}
\label{S:conc}
We have investigated the entropy for a matter core inside black holes obtained from the quantum mechanical description
of the gravitational collapse of a ball of dust  of mass $M$ introduced in Ref.~\cite{Casadio:2021cbv}.
Improving on that model, we have explicitly considered the wavefunction for the average radius $R$ of the outermost layer of dust
with mass $\mu = \epsilon\, M$, for $0 < \epsilon < 1$, which is again shown to belong to a discrete spectrum similar
to the one for the hydrogen atom.
Like in Ref.~\cite{Casadio:2021cbv}, the ground state quantum number $N_M$ is found to be proportional to $M^2$ in Planck units.
For black holes of astrophysical size, such ground states are therefore described by Laguerre polynomials of a huge 
order, which makes their analysis very difficult, both analytically and numerically. 
\par
In particular, two different types of information entropy were employed to probe this scenario, namely the DCE and the continuous
limit of Shannon's entropy, both of which can only be computed numerically and for relatively small principal quantum numbers
$\bar n=N_M+n$.
The results obtained for both entropies are qualitatively similar, showing a (much) higher information entropy for excited states
(with $n>0$) when compared to the ground state (with $n=0$). 
A higher value of the (Shannon's) entropy usually signals a larger instability.
Our results are therefore compatible with the classical dynamics of a ball of dust, which is necessarily going to collapse and
shrink under its own weight without loss of energy encoded by the ADM mass $M$.
One can then look at the sequence, from top to bottom, of black dots in Figs.~\ref{55} as representing the time evolution
of the areal radius $R(\tau)=\bar R_{\bar n}$ with decreasing values of $\bar n=N_M+n$ in correspondence with increasing
proper time $\tau$.
The existence of a ground state halts this process at a finite macroscopic size $\bar R=\bar R_{N_M}$, after which the ball
can only shrink further by losing energy $M$, so that $N_M$ decreases (hence jumping from a blue dot to the one on its
left).
It is interesting to note that this behaviour is qualitatively very different from the hydrogen atom.
In fact, an electron bound to the nucleus will ``shrink'' to smaller quantum states necessarily by emitting energy in the
form of radiation.
The dust ball instead collapses without emitting (gravitational) energy, because of the strict spherical symmetry,
until it reaches a minimum (quantum) size.~\footnote{For a more realistic (less symmetrical and with pressure)
model one expects that the collapse is instead always associated with the emission of energy.}
\par
Furthermore, we found that both kinds of entropy for the ground state show a logarithmic dependence on $M^\alpha$, with $0<\alpha<1$,
which can be extrapolated to arbitrarily  large values of $M$.
This behaviour is consistent with the notion that the number of matter microstates is proportional to the total mass (hence volume),
albeit to a non-trivial power.
As gravitational degrees of freedom are instead expected to satisfy Bekenstein's area law, the matter entropy we computed should be 
(largely) overcome by the gravitational entropy for black holes.
\par
Finally, looking at the entropy for different values of the fraction $\epsilon$, we have found that both the information entropy and the DCE
are smaller for larger fractions of dust in the outermost layer.
Since larger values of the entropy should correspond to more unstable configurations, this result seems to favour the
accumulation of matter in the outermost layer, due to quantum pressure.
The actual density profile of the final core, however, will have to be further investigated by considering refined models of the
collapsing ball with more layers.
\section*{Acknowledgments}
R.C.~is partially supported by the INFN grant FLAG and his work has also been carried out in
the framework of activities of the National Group of Mathematical Physics (GNFM, INdAM).
P.M.~thanks Coordena\c{c}\~ao de Aperfeicoamento de Pessoal de Nivel Superior Brasil 
(CAPES) -- Finance Code~001.
R.dR.~is grateful to FAPESP (Grants No. 2022/01734-7 and No.~2021/01089-1)
and the National Council for Scientific and Technological Development -- CNPq (Grant No.~303390/2019-0),
for partial financial support.  
\appendix
\section{Mass rescaling}
\label{A:rescaling}	
\setcounter{equation}{0}
The quantisation law~\eqref{N_M} makes it impractical to determine values of the mass $M$ corresponding
to integers $\bar n$ for generic values of $\epsilon$.
For studying the effect of varying $\epsilon$ on the entropies, it is therefore more convenient to rescale
the mass as
\be
\label{ap-2}
M
=
\frac{\mpl\,\tilde{M}}{\sqrt{\epsilon\left(1-\epsilon\right)}}
\ ,
\ee
so that Eq.~\eqref{ndef} is given by $\bar{n}=\tilde{M}^{2}+n$ and the wavefunctions~\eqref{radial-wavefunction}
read
\be
\label{ap-3}
\Psi_{\bar{n}}(R)
=
\sqrt{\frac{\epsilon^{3/2}\,\tilde M^9}{\pi\,\bar{n}^{5}\left(1-\epsilon\right)^{3/2}}}\,
\exp\!\left(-\frac{\epsilon^{1/2}\,\tilde M^3\,R}{\bar n\left(1-\epsilon\right)^{1/2}\lp}\right)
L_{\bar{n}-1}^{1}\!\!\left(\frac{2\,\epsilon^{1/2}\,\tilde{M}^{3}\,R}{\bar n\left(1-\epsilon\right)^{1/2}\lp}\right)
\ .
\ee
\bibliographystyle{unsrt}

\end{document}